\documentclass{mem}
\usepackage{natbib}\usepackage{txfonts}\usepackage{balance}
\usepackage{graphicx}
\usepackage[a4paper]{hyperref}

\begin{document}
\input psfig.tex

   \title{Galaxies spatially coincident with
   the JKCS\,041 X-ray emission display a red
   sequence at $\mathbf{z\sim2.2}$} 
 
   \author{S. Andreon 
          \inst{1}
          \and
          M. Huertas-Company
	  \inst{2,3}
          }

   \institute{
             Observatorio Astronomico di Brera, Milan\\
             \email{stefano.andreon@brera.inaf.it} 
	     \and
   GEPI, Paris-Meudon Observatory 5, Place Jules Janssen, 92190, Meudon, France\\
              \email{marc.huertas@obspm.fr}
         \and
             Universit\'e Paris Diderot, 75205 Paris Cedex 13, France\\
             }

\authorrunning{Andreon \& Huertas-Company}

\titlerunning{Red sequence at $z\sim2.2$}

\abstract{
New deep $z'-J$ data readly show a narrow red
sequence co-centered with, and similary concentrated to,  
the extended X--ray emission of the cluster of galaxies JKCS\,041. 
The JKCS\,041 red sequence is 
$0.32\pm0.06$ mag redder in $z'-J$ than the red sequence of the
$z_{spec}=1.62$ IRC0218A cluster, putting JKCS\,041 at $z\gg1.62$.
The colour difference of the two red sequences gives a red-sequence based
redshift of $z=2.20\pm0.11$ for JKCS\,041, where the uncertainty accounts 
for uncertainties  in stellar synthesis population models, in photometric
calibration and in the red sequence colour of
both JKCS\,041 and IRC0218A clusters. 

   }

   \maketitle

\section{Introduction}

\begin{figure*}  
\centering 
\centerline{\psfig{figure=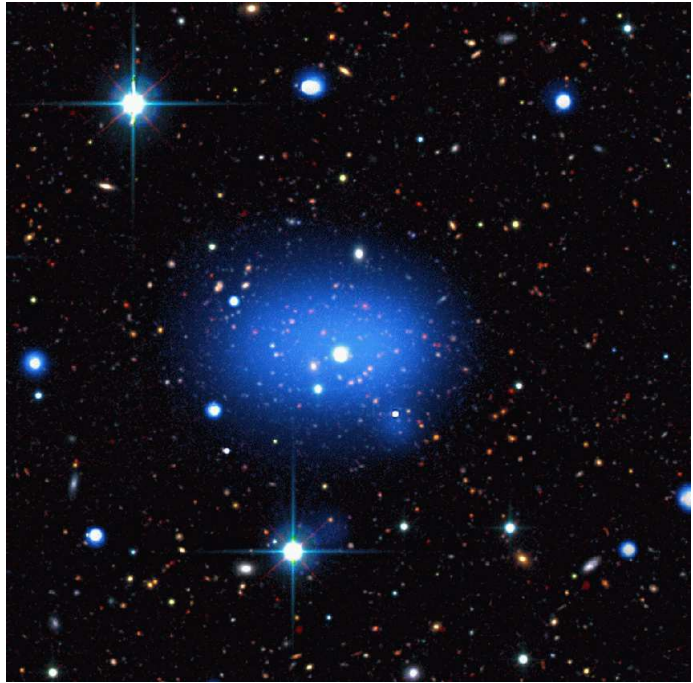,width=9truecm,clip=}}
\caption{$gz'K$-band image of JKCS\,041. Red, green and blue channel use Ks,
$z'$ and g bands. The smooth, blue emission is the X-ray emission. North
is up, East is to the left, the field of view is $5\times5$ arcmin. Reprinted
with permission.   } 
\label{fig:malaspina} 
\end{figure*}

Clusters markately differ from filaments, walls 
and sheets, up to the highest redshifts at which we have 
good data to allow such a discrimination.
These structures are very different environments in terms
of (hot) gas temperature and density, galaxy density,
velocity dispersion, morphological composition, depth of
the potential well, etc. While someone might argue that
the difference between them is just semantic, as a matter of fact,
we don't known any astronomer calling ``cluster" the Great Wall.

If one wants to study the evolution of galaxies in clusters, of the X-ray 
scaling relations of clusters, or any sensible {\it cluster} quantity, 
it seems not reasonable to build samples where clusters, filaments, 
walls and sheets enter (and exit) from the sample in an uncontrolled 
way. For example, a cluster sample contaminated at high redshift 
by the presence of 
filaments, walls and sheets is prone to confuse the 
effect of (look-back) time (i.e. evolution) with environment. Such a 
contamination is possible at high redshift if the studied sample 
includes systems that we actually don't known whether they are clusters
or not.

In order
to discriminate high redshift clusters from other large scale 
strucures (redshift spikes, filaments, sheets, etc),
a reliable probe of the structure size in the 
line of sight direction is the most difficult observation
to acquire, because measurements 
are generally easier in the plane of the sky. 
Low quality velocity dispersions, with common small samples, are of little help 
unless the found velocity dispersion is large enough, and its error 
small, to discard low values typical of large scale structures or of 
galaxy pairs. X-ray is useful, because if an X-ray source is 
unambigously extended and spatially coincident with a galaxy overdensity, 
it testifies the presence of a potential well deep enough to hot and 
retain the intracluster medium, i.e. to reject non-cluster 
possibilities. However, a generic X-ray detection, such as 
such as those of some high redshift structure candidates from 
Chiaberge et al. (2009) or Henry et al. (2010), 
is insufficient in this respect because it could originate 
from the ICM but also from the numerous (at the cluster count rate) 
point sources (or a blend of them) of the X-ray sky. 
Distinguishing between a truely extended or a blend of point sources
with just the few counts typical of faint X-ray detections
requires a good PSF, like the one of Chandra.
We emphasize that if a probability computation is used instead
of Chandra data, it should be realistic, and allow, for example,
blended point sources to have all possible fluxes (as in the
real universe) and not only a single
value as in Henry et al. (2010) computation.

Equally dangerous is allowing to enter in an uncontrolled
way structures which won't necessarily become clusters by $z=0$.
Several proto-clusters are in this situation, basically because
the 1 (or 2) $\sigma$ interval of the collapsing time includes
values larger than the look-back time of the structure. For
example, the ''proto-clusters" in Hatch et al. (2010) have a collapsing
time of 6 Gyr, shorter than the look back time at the redshift of the
systems (about 10.8 Gyr), but widely uncertain and larger that the latter 
at better than 1.5 sigma. 
Therefore, there is very little, if any, evidence that these systems will be clusters by 
today. Assuming that they will become clusters by $z=0$, when such evidence
is lacking, is risky and prone to mix different environments at different
redshifts.

To summarize, having detected a structure at high redshift, it not enough
to call it cluster. If the nature of the detected structure is
unknown, we risk to compare apples (at very high redshift) to oranges 
(at lower redshift) and therefore call ``evolution" what
is instead ``contamination" or ``environemnt".

In that sense, JKCS\,041 (Andreon et al. 2009, see Fig 1) is a
uniq very high redshift cluster, as it has
an unambiguously extended X-ray emission (from Chandra data)  
testifying the presence of an intracluster medium and deep
potential well, that makes it unambigously a cluster. However,
it lacks of a spectroscopic redshift: its 68 \%
photometric redshifts  
interval  [1.89,2.12]. In this proceeding
we first report the presence of a clear red sequence of
passive galaxies in the region of JKCS\,041 using a filter pair 
sampling the Balmer break at $z>1.2$: the $z'-J$ colour. 
Second, by comparing the red sequence colour of the JKCS\,041 
and IRC0218A, a spectroscopically confirmed cluster at $z\sim1.62$ 
(Papovich et al. 2010: Tanaka et al. 2010), 
we confirm that JKCS\,041 lies at much
higher redshift,
and we quantify its photometric redshift ($z\sim 2.2$). More details
can be found in Andreon \& Huertas-Company (2010).

Throughout the
paper, we assume the following cosmological parameters:
$H_0=70$ km s$^{-1}$ Mpc$^{-1}$, $\Omega_m=0.3$ and
$\Omega_\Lambda=0.7$. Magnitudes are in the AB system. 

\begin{figure*}
\centerline{\psfig{figure=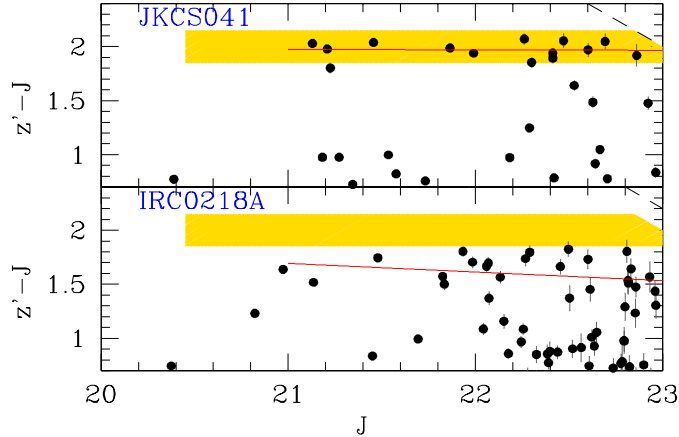,width=9truecm,clip=}}
\caption{Colour--magnitude plot in the direction of JKCS\,041 (top panel) 
and toward IRCS0218A (bottom panel). 
Solid lines shows the fitted colour-magnitude relations of
JKCS\,041 (top panel) and IRC0218 (bottom panel) clusters, 
whereas shading marks the simpler colour range
$1.85<z'-J<2.15$. Slanted 
dashed lines mark magnitudes where the S/N=15 in 
$z'$, and therefore delimit the region where catalogs should be
complete. JKCS\,041 red sequence 
galaxies are 0.32 mag redder than the red sequence of IRC0218A, at $z=1.62$,
indicating that JKCS\,041 is at $z\gg1.62$.
} 
\label{fig:CMRs}
\end{figure*}

\section{Data and minor photometric corrections} \label{sec:dataset} 

JKCS\,041 (see Fig. 1) is in the area covered by  
CFHTLS deep survey release 
and by WIRDS follow-up in the infrared filters
($J, K$) (Bielby et al.,
in preparation, catalogs are available on the Terapix site). 
More precisely, we use the catalog generated using
$K$-band as detection image and the other bands ($z'$ and $J$)
in analysing mode.

IRC0218A is 2.1 degree aways from JKCS\,041 and is in the area covered by  
Williams et al. (2009), whose catalogs are based on
UKIDSS survey (Lawrence et al. 2007) for
the infrared bands ($J$ and $K$), and from the
SXDS survey (Furusawa 2008) 
for the $z'$ band. Colors listed in the catalog are 
measured in matched apertures and corrected for seeing
differences.

The photometry of the two catalogs comes from different telescopes 
and reduction pipelines. The comparison of 
the stellar locii in the $z'-J$ vs $J-K$ plane confirms
that the two catalogs are truely on a consistent
photometric system, after a minor color correction, 
$0.039 \pm 0.014$ mag, is applied to JKCS\,041 $z'-J$
colors.

\begin{figure*}
\centerline{\psfig{figure=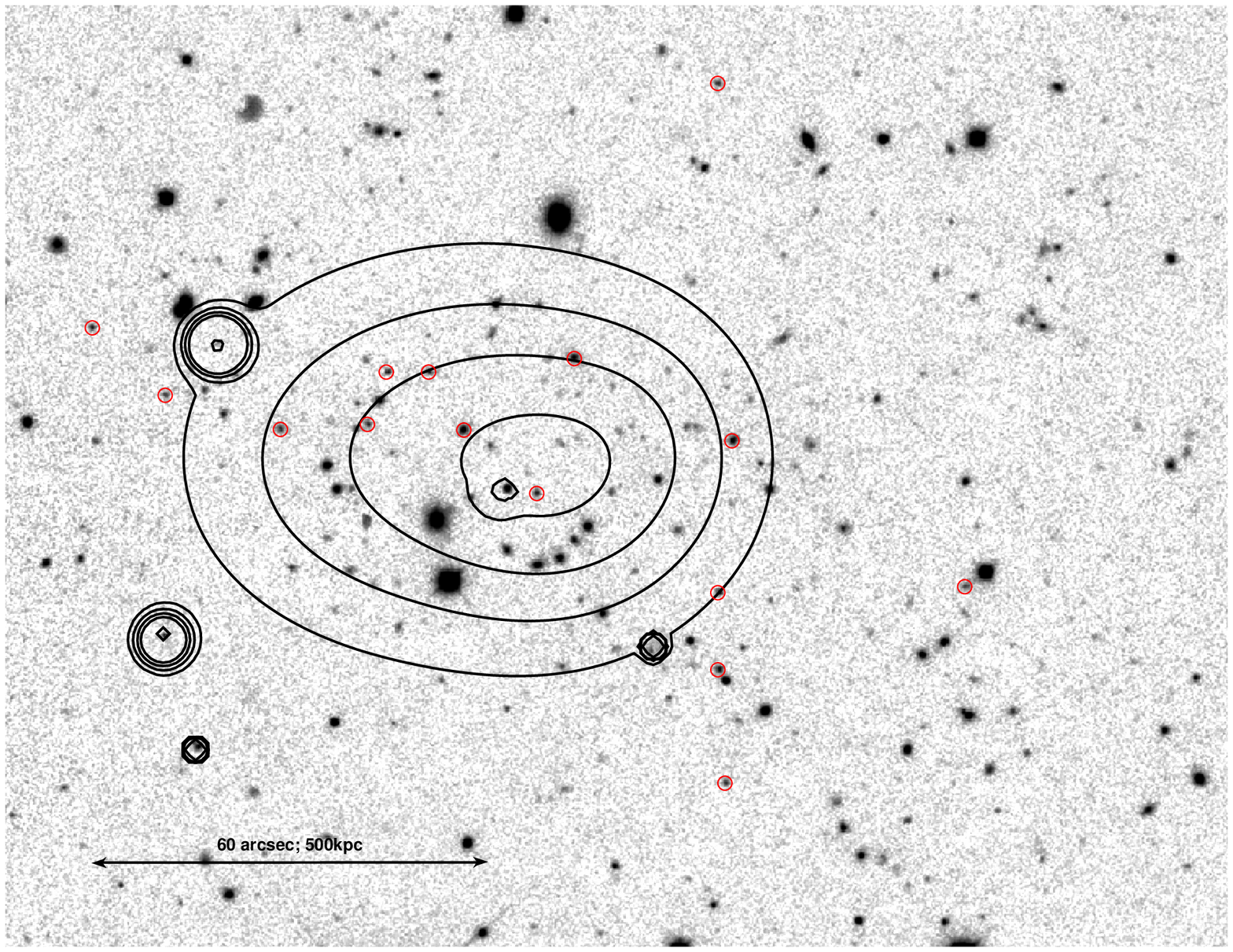,width=9truecm,clip=}}
\caption{$J$ band image of the field near to JKCS\,041.
Contours mark the adaptively smoothed X--ray emission detected by Chandra 
(from Andreon et al. 2009). Circles mark galaxies on the red sequence
$1.85<z'-J<2.15$. These are concentrated and co-centered with 
the X--ray emission. North is up and East is to the left. The ruler
is 1 arcmin wide ($\sim 500$ kpc at $z=2.2$).
} 
\label{fig:boh}
\end{figure*}

\section{Results} 

\subsection{Red sequences \& JKCS\,041 photometric redshift} 

The top panel of  Figure~2
shows the color-magnitude relations toward JKCS\,041 within a
radius of $1$ arcmin ($\sim 0.5$ Mpc at $z\sim2$), 
after applying the minor photometric correction
described above and a partial masking of 45 deg sector,
partially contaminated by another structure 1.2 arcmin away. 
The bottom panel 
shows for comparison
a random control region of the same solid angle (it turns
out to be at roughly 0.4 deg north of JKCS\,041). Slanted and vertical
dashed lines mark magnitudes where the S/N=15 in $J$ (vertical)
or $z'$ (diagonal), and therefore delimit the region where catalogs should be
complete.
JKCS\,041 presents a clear red sequence of $\sim14$
objects well aligned, within a narrow $\sim 0.1$ colour band, all within
1 arcmin from the X-ray cluster center. We fit the red sequence through the Bayesian
methods of Andreon (2006) and Andreon et al. (2006), also used for
other clusters (Andreon 2008; Andreon et al. 2008), solving
at once for all parameters: slope, intercept and intrinsic spread of the 
colour-magnitude relation, luminosity function parameters (characteristic
magnitude, faint end slope $\alpha$ and normalisation),
accounting for photometric errors and also the presence of
a fore/background population, the latter constrained using
a control field region with a $64$ times larger solid angle. 
The mean fitted red sequence is marked in
the figure as solid line, at $z'-J\sim 2$ mag. 

Fig. 3 shows the spatial
distribution of galaxies within the shaded colour range, $1.85<z'-J<2.15$, with 
no spatial filtering applied: these galaxies are not uniformly distributed
but concentrate in the region of the X-ray emission.

We also fitted the color-magnitude of IRC0218A.
The solid, slanted, line at $z'-J\sim 1.6$ mag in the bottom panel of
Fig. 2 marks it.
We found the
IRC0218A red sequence 
$0.32\pm0.06$ mag bluer than the JKCS\,041 red sequence. The quoted
error accounts for the error in the two red sequence intercepts 
($0.035$ and $0.05$ mag for JKCS\,041 and IRC0218A, respectively) and
the uncertainty in the photometric correction ($0.014$ mag, see sec 2).
The IRC0218A reddest galaxies are all bluer than the bluest galaxies on the
JKCS\,041 red sequence, and in fact none fall in the shaded area where
JKCS\,041 red sequence galaxies are found. 
The redder colour of the JKCS\,041 red sequence directly implies
that $z\gg 1.62$. Even without any photometric correction,
the JKCS\,041 red sequence is 0.29 mag redder than the IRC0218A red sequence,
showing the much larger redshift of JKCS\,041 compared to $z=1.62$.

\begin{figure}
\centerline{\psfig{figure=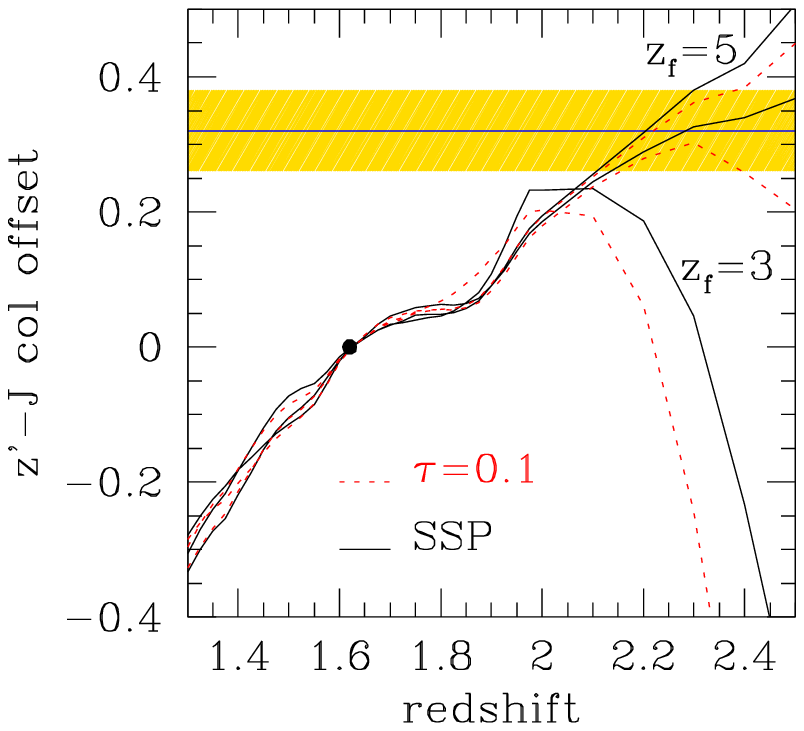,width=6truecm,clip=}}
\caption{Photometric estimate of the JKCS\,041 redshift. Colour 
track of SSP (solid black) and $\tau=0.1$ (red dotted)
models with different formation
redshifts, $z_f=5,4,3$ (from top to bottom), are plotted, after
zero-pointing them to the observed colour of IRC0218A cluster
at $z=1.62$. The measured colour difference is marked with an horizontal
line, and its error shaded. By simple eye inspection of this figure,
the JKCS\,041 photometric redshift is $z\sim 2.2$. 
} 
\label{fig:photoz}
\end{figure}

Fig. 4 plots colour 
tracks of single stellar populations (SSP, solid black) and  
exponential declining ($\tau=0.1$, red dotted) models with solar
metallicity, Chabrier initial mass function, formed at $z_f=5,4,3$, 
zero-pointing them to the observed colour of IRC0218A cluster
at $z=1.62$.  This approach to estimate the
photometric redshift is more robust 
than using the absolute colour of
the population, and benefits of a smaller extrapolation. 
For computing the colour tracks we use the 2007 version of
Bruzual \& Charlot (2003) synthesis population model.
The measured colour difference is marked with an horizontal
line, and its error is shaded. The JKCS\,041 photometric redshift derived
from a SSP model with
$z_f=5$ can be easely read in it:
$z_{phot}=2.20 \pm 0.10$. We add in quadrature a further 0.05 error term
to approximatively account for different $z_f$ and star formation histories
(this accounts for the $0.07$ total spread in redshift of the
three models reaching a colour difference of $0.32$ mag)
giving a final photometric redshift of $z_{phot}=2.20 \pm 0.11$.

The newly determined photometric redshift is slightly higher than the
conservatively estimated photometric redshift quoted
in Andreon et al. (2009)
and consistent with it, since the two 68 \% confidence intervals 
([1.84,2.12] vs [2.09,2.31]) overlap. The current redshift determination 
superseeds 
the old determination, since we now use a calibration at similar
redshift (via IRC0218 red sequence colour).

Bielby et al. (2010)
detected, using the very same photometric catalog used here,
a cluster co-centered with the X-ray cluster emission and 
formed by much the very same galaxies plotted in the top panel of Fig 2.
They put their detection at $z_{phot}=1.39$ which is in flagrant disagreement
with the observed colour of the red sequence of the spectroscopically
confirmed IRC0218 cluster: a $z=1.39$ red sequence should be 0.2
mag bluer than the IRC0218 red sequence (Fig. 4), whereas the observed 
red sequence is 0.32 mag redder.

The abundant spectroscopic (VVDS, Le Fevre et al. 2005) 
and photometric ($>12$ bands) data available
for galaxies in the X--ray area of JKCS041 has attracted many authors to look
for clusters in this region using a variety of cluster detection methods
(e.g. Olsen et al. 2007, 2008; Grove et al. 2009; Mazure
et al. 2007; Thanjavur et al. 2009). However, up to now, no other cluster
co-centered with the X-ray emission has been found even, if as showed in this
work, the quality of available data enables a detection at $z\sim 2.2$. 
Therefore, the galaxy cluster at $z\sim 2.2$ is the only 
cluster from which the extended X--ray emission may come from.

\section{Conclusions}

We show that galaxies $0.32$ mag redder than the red sequence of
the $z_{spec}=1.62$ cluster IRC0218A are spatially concentrated
where the JKCS\,041 X-ray emission is located. Their red colour
implies that the cluster JKCS\,041 is at $z=2.20\pm0.11$, 
where the uncertainty accounts 
for uncertainties  in stellar synthesis population models, in photometric
calibration and in the red sequence colour of
both JKCS\,041 and IRC0218A clusters.

We can hence confirm that JKCS\,041 is a cluster of galaxies with 
the photometric redshift $z_{red \ sequence}=2.20\pm0.11$, with a formed
potential well, deep enough to hot and retain the intracluster medium,
and with a well defined red sequence. Incoming X--ray survey telescopes
or red-sequence based surveys
will likely return hundreds of $z\sim2$ clusters candidates.
Getting spectroscopic redshifts for all of them, or even a small part, 
is too time consuming with present telescopes. Therefore,
photometric redshifts based on the red sequence color will necessarily become
very popular in the next years and we need to get used to them.  
Sunayev-Zeldovich surveys (e.g. High et al. 2010 for SPT; Menanteau et al.
2010 for ACT) are already using galaxy colours to infer
the cluster redshift for their small
(dozen of clusters) samples at low ($z<1$) redshift.

\begin{acknowledgements}
We thank Marco Malaspina for Fig. 1 compositing. 
Based on observations obtained with
MegaPrime/MegaCam\footnote{The full text acknowledgement is at
http://www.cfht.hawaii.edu/Science/CFHLS/cfhtlspublitext.html} and 
WIRCAM\footnote{The full text acknowledgement is  at
http://ftp.cfht.hawaii.edu/Instruments/Imaging/WIRCam/WIRCamAcknowledgment.html}
at CFHT. 
\end{acknowledgements}

{}

\end{document}